\documentclass{aa}
\usepackage{psfig,latexsym,longtable,lscape}

\def\it{\sl}
\def\asec{\ifmmode ^{\prime\prime}\else$^{\prime\prime}$\fi}
\def\amin{\ifmmode ^{\prime}\else$^{\prime}$\fi}

\newcommand\approxgt{\mbox{$^{>}\hspace{-0.24cm}_{\sim}$}}

\begin{document}

   \thesaurus{08         
               08.02.1;  
               13.25.5;  
               08.05.3;  
               08.23.1;  
               11.13.1)) 
             } 
   \title{Luminous supersoft X-ray emission from the recurrent nova U~Scorpii}

\author{P. Kahabka\inst{1} \and H.W. Hartmann\inst{2} \and A.N. Parmar\inst{3} 
           \and  I. Negueruela\inst{4}
          }

\offprints{ptk@astro.uva.nl}
 
\institute{$^1$~Astronomical Institute and Center for High Energy
Astrophysics, University of Amsterdam, Kruislaan 403, NL-1098~SJ 
Amsterdam, The Netherlands\\
$^2$~SRON Laboratory for Space Research, Sorbonnelaan 2, 3584 CA Utrecht,
The Netherlands\\
$^3$~Astrophysics Division, Space Science Department of ESA, ESTEC,
P.O. Box 299, 2200 AG Noordwijk, The Netherlands\\
$^4$ SAX Science Data Center, ASI, c/o Telespazio, via Corcolle 19, 
I00131 Roma, Italy}

\date{Received 7 May 1999 / Accepted 15 June 1999}
 
\maketitle
\markboth{P. Kahabka et al. Luminous supersoft emission from U~Sco}{}
 
\begin{abstract}

BeppoSAX detected luminous 0.2--2.0~keV supersoft X-ray emission from the 
recurrent nova U~Sco $\sim$19-20 days after the peak of the optical outburst 
in February 1999. U~Sco is the first recurrent nova to be observed during a 
luminous supersoft X-ray phase. Non-LTE white dwarf atmosphere spectral models
(together with a $\sim$0.5~keV optically thin thermal component) were fitted 
to the BeppoSAX spectrum. We find that the fit is acceptable assuming
enriched He and an enhanced N/C ratio. This implies that the CNO cycle was 
active during the outburst, in agreement with a thermonuclear runaway scenario.
The best-fit temperature is $\rm \sim9\times 10^5\ K$ and the bolometric 
luminosity $\rm (0.16-1.2)\times 10^{36}\ erg\ s^{-1}\ (d/kpc)^2$. These 
values are in agreement with those predicted for steady nuclear burning on a 
WD close to the Chandrasekhar mass. The fact that U~Sco was detected as a 
supersoft X-ray source is consistent with steady nuclear burning continuing 
for at least one month after the outburst. This means that only a fraction of 
the previously accreted H and He was ejected during the outburst and that the 
WD can grow in mass, ultimately reaching the Chandrasekhar limit. This makes 
U~Sco a candidate type Ia supernova progenitor.

 
   \end{abstract}
%

\section{Introduction}
Recurrent novae (RN) are a small and diverse subclass of cataclysmic variables,
which show multiple outbursts resembling those of classical novae, though of 
lesser magnitude (see Webbink et al. 1987; Sekiguchi 1995). U~Scorpii (U~Sco) 
is one of the six best known members of this class. The source underwent 
outbursts in 1863, 1906, 1936, 1979, 1987, and most recently on 1999 Feb 25.2 
(Schmeer et al. 1999). The last two outbursts were separated by 8 and 12 years
respectively. It is the RN with the shortest recurrence period known. 

Starrfield et al. (1988) applied thermonuclear runaway (TR) theory to this nova
assuming a very massive white dwarf (WD).

The estimate of the distance to U~Sco in the literature varies with different
assumptions. Kato (1990) obtained a distance range 3.3--8.6 kpc comparing the
observed visual light curve with the theoretical one and assuming a high mass
($\rm \sim1.38\ M_{\odot}$) WD. If the donor star is a dwarf the distance can
be $\rm 3.5\pm 1.5~kpc$ (Hanes 1985), however most authors recently agree
instead on a {\it subgiant} nature of the donor, as indicated by the detection
of a Mg~Ib absorption feature at $\lambda$5180 in the late 1979 outburst 
spectrum, consistent with a K2~III spectral type (Pritchet et al. 1977). 
This is in agreement with a low mass ($\simeq$ 1 M$_\odot$) subgiant secondary
with M$_V$=+3.8 which fills the Roche lobe at an orbital period 
$\simeq$1.2 days (Portegies Zwart, private communication). From the apparent  
magnitude V=20.0 in the faint state and the visual extinction A$_V$=0.6 a 
distance $\simeq$13~kpc is derived, in rough agreement with d=14.8 kpc derived
by Webbink et al. (1987) with the assumption of a G~III subgiant, and with 
d$\simeq$14~kpc estimated by Warner (1995). In any case U Sco is at a latitude
21$^o$ and for any distance d$\geq$ 3~kpc it belongs to the galactic halo 
population. It is seen through the full galactic hydrogen column of 
$\rm 1.4\times 10^{21}\ cm^{-2}$ (Dickey \& Lockman 1990). 

U~Sco was observed to be an eclipsing system by Schaefer (1990). The orbital 
period is 1.23~days (Johnston \& Kulkarni 1992; Schaefer \& Ringwald 1995). 
$\rm m_V$ varies from 18.5 to 20. An accretion disk is required from the 
modeling of the optical continuum during quiescence. The maximum visual 
magnitude during outburst is $\rm m_V\sim8$. U~Sco shows the fastest visual 
decline of 0.67 magnitude per day of all known novae (Sekiguchi et al. 1988). 
Spectroscopically it shows very high ejection velocities of 
$\rm \sim(7.5-11)\times 10^3\ km\ s^{-1}$ (Williams et al. 1981; Rosino \& 
Iijima 1988; Niedzielski et al. 1999). 

Ejecta abundances have been estimated from optical and UV studies (Williams 
et al. 1981; Barlow et al. 1981). From the emission lines a depletion in 
hydrogen relative to helium with He/H$\sim$2 has been derived while the CNO
abundance was solar with an enhanced N/C ratio. The strongest emission feature 
at maximum is the He{\sc ii} $\lambda4686$ line. Other reported lines are 
H$\alpha$, He{\sc i}, He{\sc ii}, N{\sc ii}, N{\sc iii}, C{\sc iii}, and 
C{\sc iv} (Zwitter et al. 1999; Bonifacio et al. 1999). Satellite lines to 
H$\alpha$, He{\sc ii} and He{\sc i} have also been detected (Bonifacio et al. 
1999). The estimated mass of the ejected shell for the 1979 outburst is 
$\rm M_{shell}\sim10^{-7}\ M_{\odot}$ (Williams et al. 1981). 

It has been suggested that the companion of U~Sco may be somewhat evolved (and
helium enriched) as the quiescent spectrum shows strong He{\sc ii} emission 
lines (cf. Hanes 1985). Hachisu et al. (1999) propose an evolutionary scenario
for this system assuming a secondary star which experienced a helium accretion
phase. The WD may efficiently grow in mass towards the Chandrasekhar (CH) 
limit and explode as a SN~Ia (Della Valle \& Livio 1996). But if U~Sco is in 
the galactic halo then the system belongs to an old stellar population and the
evolution may be different. Helium enrichment as observed from U~Sco may also 
be due to helium enriched winds from the WD (cf. Prialnik \& Livio, 1995).

\section{The BeppoSAX observation}

After the optical outburst of U~Sco was reported (Schmeer et al. 1999) a 
target of opportunity observation of U~Sco was performed with the BeppoSAX 
X-ray satellite. According to the calculations of Kato (1996), supersoft 
(SSS) X-ray emission is predicted to be observed $\sim$10--60~days after the 
optical outburst. The 50~ks exposure observation was performed during 1999 
March 16.214 -- 17.425, 19--20~days after the optical outburst. Here we 
report the first results of an analysis of the mean X-ray spectrum observed 
during this observation.

The scientific payload of BeppoSAX (see Boella et al. 1997a) comprises four 
coaligned Narrow Field Instruments including the LECS (Parmar et al. 1997) and 
MECS (Boella et al. 1997b). U~Sco was detected with mean LECS and MECS net 
count rates, after background subtraction, of $\rm (5.67\pm0.23)\times 
10^{-2}\ s^{-1}$ and $\rm (1.35\pm0.38)\times 10^{-3}\ s^{-1}$ respectively. 
The source was not detected in the high-energy non-imaging instruments. The 
X-ray flux varies by a factor of $\sim$1.5 during the observation possibly 
due to orbital variations or a rise in flux.   

\begin{table}
\caption[]{Best-fit values derived from spectral fits to the BeppoSAX LECS and
MECS spectrum of U~Sco using (a) a blackbody model with absorption edges and 
(b) an optically thick non-LTE WD atmosphere model (with He enriched and the 
N/C ratio enhanced) and an optically thin Raymond and Smith component 
(assuming He enriched and the ratio N/C enhanced). 90\% confidence parameter 
ranges are given. For the edges the absorption depth at the given energies are
listed}
      \begin{flushleft}
      \begin{tabular}{llc}
      \hline
      \noalign{\smallskip}
\multicolumn{3}{c}{(a) Blackbody with absorption edges}                     \\
      \noalign{\smallskip}
      \hline
      \noalign{\smallskip}
 Parameter          & unit                           & 90\% confidence      \\ 
      \noalign{\smallskip}
 $\rm N_{H}$        & ($\rm 10^{21}\ cm^{-2}$)       & 1.8 - 2.6            \\
 $\rm T$            & ($\rm eV$)                     & 102 - 112            \\
                    & ($\rm 10^6\ K$)                & 1.2 - 1.30           \\
 $\rm R$            & ($\rm 10^7\ cm\ (d/kpc)$)      & 0.50 - 0.66          \\
 $\rm L$ & ($\rm 10^{35}\ erg\ s^{-1}\ (d/kpc)^2$) & 0.39 - 0.88            \\
 \noalign{\smallskip} 
 \multicolumn{3}{c}{Absorption edges}                                       \\
 \noalign{\smallskip} 
 $\rm \tau$ N~{VI}  & 0.55~keV                     & 3.8 - 5.6              \\
 $\rm \tau$ N~{VII} & 0.67~keV                     & 1.6 - 3.7              \\
 $\rm \tau$ O~{VII} & 0.74~keV                     & $<$1.6                 \\
 $\rm \tau$ O~{VIII}& 0.87~keV                     & 4.5 - 7.2              \\
      \noalign{\smallskip}
      \hline
      \noalign{\smallskip}
\multicolumn{3}{c}{(b) WD atmosphere with He enriched and ratio N/C enhanced}\\
      \noalign{\smallskip}
      \hline
      \noalign{\smallskip}
 $\rm N_{H}$        & ($\rm 10^{21}\ cm^{-2}$)       & 3.1 -- 4.8           \\
 $\rm T$            & ($\rm eV$)                     & 73.7 -- 76.3         \\
                    & ($\rm 10^5\ K$)                & 8.5 -- 8.9           \\
 $\rm R$            & ($\rm 10^7\ cm\ (d/kpc)$)      & 1.9 -- 5.5           \\
 $\rm L$ & ($\rm 10^{35}\ erg\ s^{-1}\ (d/kpc)^2$)   & 1.6 -- 12            \\
 \multicolumn{3}{c}{Raymond-Smith component}                                \\
 \noalign{\smallskip} 
 $\rm kT$           & ($\rm keV$)                    & 0.22 -- 0.52         \\
 $\rm EM$ & ($\rm 10^{55}\ cm^{-3}\ (d/kpc)^2$)      & 0.44 -- 3.2          \\
 \noalign{\smallskip} 
      \hline
      \end{tabular}
      \end{flushleft}
   \end{table}

The combined LECS and MECS spectrum was first fit with a simple blackbody 
spectral model. The fit is unacceptable with a $\chi^2$ of 72 for 10 
degrees of freedom (dof). We then added absorption edges due to highly 
ionized species of N and O expected in the hot atmosphere of a steadily 
nuclear burning WD. The edge energies were fixed at 0.55~keV, 0.67~keV, 
0.74~keV, and 0.87~keV, corresponding to the Lyman edges of N~{\sc vi}, 
N~{\sc vii}, O~{\sc vii}, and O~{\sc viii}, respectively. Only the 
N{\sc vi}, N{\sc vii}, and O{\sc viii} edges were detected at high 
significance with absorption depths of 4.3, 2.4, and 5.6, respectively. 
The O{\sc vii} edge is not detected and the 90\% confidence upper limit 
to its absorption depth is $<$1.6. The $\chi ^2$ is 12 for 6 dof. However, 
other interpretations of the spectral shape above $\sim$0.8~keV appear 
to be more likely (see below and the discussion). We independently fitted 
the edge energies of the N~{\sc vi} and N~{\sc vii} features and derived 
90\% confidence ranges of 0.524--0.555~keV and 0.630--0.669~keV, 
respectively and an absorbing hydrogen column density 
$\rm (1.8-2.6)\times 10^{21}\ atom~cm^{-2}$. 

\begin{figure*}[htbp]
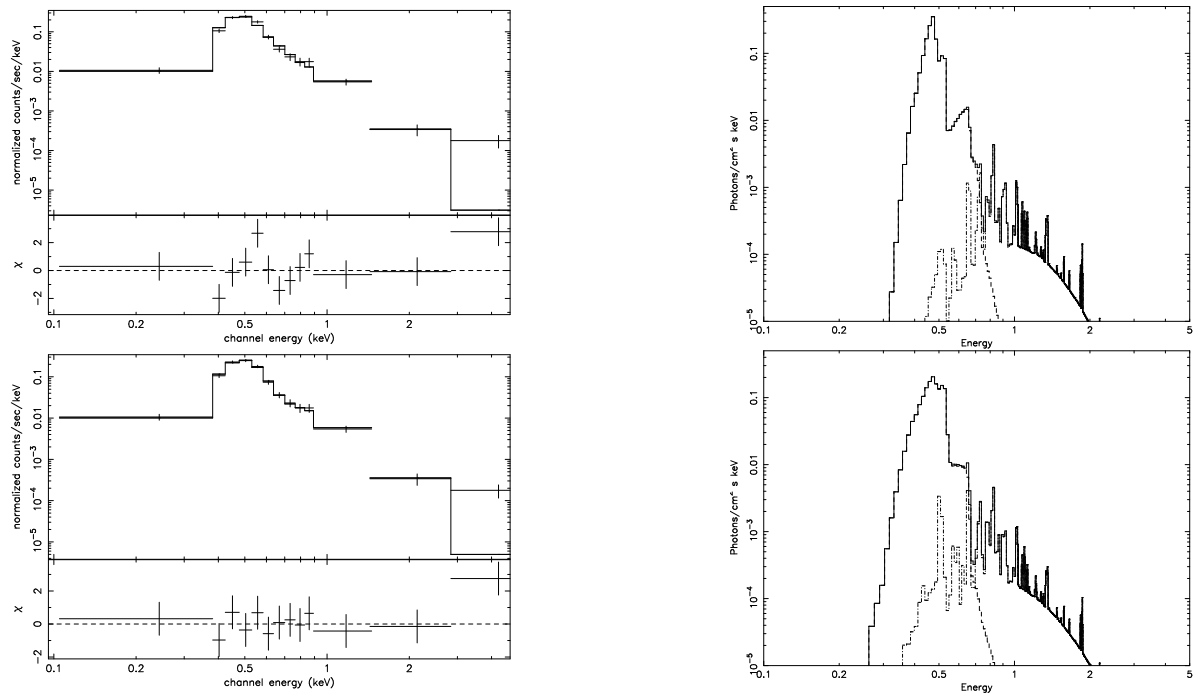

  \centering{ 
  \vbox{\psfig{figure=be074.f1,width=8.2cm,%
  bbllx=-3.0cm,bblly=1.0cm,bburx=25.5cm,bbury=16.7cm,clip=}}\par
  \vspace{-4.6cm}
  \hfill\parbox[b]{1.0cm}{}
  \vbox{\psfig{figure=be074.f2,width=7.3cm,%
  bbllx=1.0cm,bblly=0.5cm,bburx=28.5cm,bbury=17.9cm,clip=}}\par
  \vbox{\psfig{figure=be074.f3,width=8.2cm,%
  bbllx=-3.0cm,bblly=1.0cm,bburx=25.5cm,bbury=16.7cm,clip=}}\par
  \vspace{-4.6cm}
  \hfill\parbox[b]{1.0cm}{}
  \vbox{\psfig{figure=be074.f4,width=7.3cm,%
  bbllx=1.0cm,bblly=0.5cm,bburx=28.5cm,bbury=17.9cm,clip=}}\par
            }
  \caption[]{Combined BeppoSAX LECS and MECS spectra of U~Sco (left) and 
             spectral models (right). Upper panels show the non-LTE WD 
             atmosphere model using cosmic abundances, lower panels show 
             the WD atmosphere model using He enrichment and enhanced N/C 
             ratio} 
\end{figure*}

WD atmosphere spectra have been shown to deviate strongly from simple 
blackbodies (e.g., Hartmann \& Heise 1997). The use of sophisticated WD 
atmosphere model spectra is required. We applied a non-LTE WD atmosphere 
spectral model grid assuming a very massive (log(g)=9.5) WD with cosmic CNO 
abundances (see e.g., Hartmann et al. 1999). The fit was unacceptable 
at energies $\approxgt$0.8~keV. We added an optically thin thermal component 
(Raymond \& Smith 1977), hereafter RS, to the model. Such a component may 
be due to a strong wind from the WD atmosphere and has been observed in the 
classical nova Cyg 1992 (Balman et al. 1998). The fit was still unacceptable 
with a $\chi^2$ of 23.4 for 8 dof. We also fitted the observed spectrum with 
two optically thin RS components. We found that the fit was not acceptable 
with a $\chi^2$ of 55.8 for 8 dof.

When the CNO cycle is active then the N/C and O/C ratios are strongly 
modified. A strong enrichment of N with respect to C is expected as N is 
involved in the slowest reaction. We calculated log~g=9.5 non-LTE WD 
atmosphere spectral models with He and CNO (number) abundances ($\rm H=0.5$, 
$\rm C=9\times 10^{-4}$, $\rm N=6\times 10^{-3}$, $\rm O=3\times 10^{-3}$ 
with respect to helium) according to values determined from optical/UV 
studies of the nova 
ejecta of U~Sco (Williams et al. 1981). In addition, we applied a hot 
optically thin thermal component. We found that with these assumptions 
the fit was acceptable with a $\chi^2$ of 10.7 for 8 dof. The best-fit 
atmospheric temperature is $\rm (8.53 - 8.85)\times 10^5\ K$ 
(90\% confidence), the atmospheric radius is $\rm (1.9 - 5.5)\times 
10^7\ cm\ (d/kpc)$, and the bolometric luminosity $\rm (0.16 - 1.2)\times 
10^{36}\ erg\ s^{-1}\ (d/kpc)^2$. For the optically thin component we 
derive a temperature, kT, of 0.22--0.52~keV and an emission measure, EM, 
of $\rm (0.42 - 3.2)\times 10^{55}\ cm^{-3}\ (d/kpc)^2$ assuming that He 
is enriched and N/C enhanced. The absorbing hydrogen column density is 
$\rm (3.1-4.8)\times 10^{21}\ atom~cm^{-2}$. This value is larger than the 
galactic absorption in the direction of U~Sco of $\rm 1.4\times 10^{21}\ 
cm^{-2}$ (see Introduction) indicating a substantial intrinsic absorption.

\section{Discussion}
\subsection{Supersoft X-ray emission}

U~Sco belongs now to those novae for which SSS X-ray emission has been 
discovered (cf. Orio \& Greiner 1999).

The TR theory predicts two processes which can generate soft X-rays: Shock 
acceleration in the nova ejecta and steady nuclear burning. For Nova Cyg 1992
an optically thin component due to shocks has been detected. In addition an
optically thick SSS X-ray component has been observed $\sim$60~days after the 
outburst and for $\sim$600~days (Krautter et al. 1996; Balman et al. 1998). 

According to the calculations of Kato (1996) performed for U~Sco, and assuming
a massive ($\rm M_{WD}$=$\rm 1.377 M_{\odot}$) WD a SSS component is predicted
to be observed from $\sim$10~days after the outburst. In the case of the 
H-rich model (He/H=0.1) the supersoft component is expected to rise till 
$\sim$50~days after the outburst to a maximum luminosity of 
$\rm \sim3\times 10^{36}\ erg\ s^{-1}\ (d/kpc)^2$. In the He-rich model 
(He/H=2), a maximum luminosity for the SSS component of 
$\rm \sim(0.8-1)\times 10^{36}\ erg\ s^{-1}\ (d/kpc)^2$ is reached 
$\sim$20 days after the optical outburst. Using the He enriched fit with 
the N/C ratio enhanced (Table~1), the observed bolometric luminosity 
$\sim$19--20~days after the optical outburst is $\rm \sim(0.16-1.2)\times 
10^{36}\ erg\ s^{-1}\ (d/kpc)^2$. Assuming a distance $\simeq$14~kpc 
(see Introduction) a bolometric luminosity of $\rm \sim(0.3-2.4)\times 10^{38}\
erg\ s^{-1}$ is derived. The luminosity is in agreement with the bolometric 
luminosity of $\rm \sim10^{38}\ erg\ s^{-1}$ predicted for novae (Mc Donald 
et al. 1985). The temperature of $\rm (8.7-8.9)\times 10^5\ K$ and the 
luminosity of $\rm \le 2.4\times 10^{38}\ erg\ s^{-1}$ derived from the X-ray 
spectral fit requires a very massive $\rm M_{WD} > 1.2\ M_{\odot}$ WD 
consistent with an almost CH mass WD (e.g. Kato 1997).

\subsection{Spectrally hard component}

In addition to the optically thick SSS X-ray model spectrum, the spectral 
fits require a spectrally hard component. A similar component in addition to 
a SSS component was used by Balman et al. (1998) for X-ray spectral fits to 
the classical nova Cyg~1992. Using an optically thin thermal model we derive 
a temperature of $\rm 0.22-0.52~keV$, an emission measure 
$\rm EM = (0.4-3.2)\times 10^{55}\ cm^{-3}\ (d/kpc)^2$ if He is enriched and 
the ratio N/C enhanced.

If we assume a terminal wind velocity of $\rm v_{\infty} \approx 10^8\ cm\ 
s^{-1}$ the wind mass loss rate for a He/H=2 mixture can be estimated from
$\rm \dot{M}\approx14.5\ m_H\ v_{\infty}\ \sqrt{EM\ r}$. Here $r$ is the 
typical radius of the emitting region. We assume $\rm r=10^{11}\ cm$, the 
radius of the Roche-lobe, and use the result of the spectral fit assuming He 
is enriched. We then obtain a wind mass loss rate of 
$\rm \dot{M}\ =\ (2.4-6.9)\times 10^{-8}\ M_{\odot}\ yr^{-1}\ (d/kpc)$. For 
a distance to U Sco of 14~kpc, we derive a wind mass loss rate of 
$\rm (3.4-9.7)\times 10^{-7}\ M_{\odot}\ yr^{-1}$. A near-CH mass WD at 
$\rm T = 9\times 10^5~K$ experiences an envelope mass loss of 
$\rm \sim1.2\times 10^{-6}\ M_{\odot}\ yr^{-1}$ due to both steady nuclear 
burning and a wind from the WD. The steady nuclear burning mass loss can be 
estimated to be $\rm \sim6\times 10^{-7}\ M_{\odot}\ yr^{-1}$ (Hachisu et al. 
1999). The mass loss due to the wind is $\rm \sim6\times 10^{-7}\ M_{\odot}\ 
yr^{-1}$. This value is in agreement with the range we derived above. For a 
duration of the steady nuclear burning phase plus wind mass loss phase of 
$\sim$0.1~year (Kato 1996) we derive a mass loss from the WD envelope of 
$\rm \sim1.2\times 10^{-7}\ M_{\odot}$ of which $\rm \sim6\times 10^{-8}\ 
M_{\odot}$ is due to the wind. In addition the predicted post-outburst 
envelope mass is $\rm \sim8\times 10^{-8}\ M_{\odot}$ (Hachisu et al. 1999). 
This would mean that 70\% of the envelope mass has remained on the WD allowing
it to increase in mass. Williams et al. (1981) derive from the UV lines (for 
14~kpc and $\rm r > 6.5\times 10^{11}\ cm$) a wind mass loss rate 
$\rm \dot{M}\ > 5.4\times 10^{-8}\ M_{\odot}\ yr^{-1}$ which differs 
significantly from our value, although it is subject to many uncertainties, 
and differences between outbursts cannot be accounted for.

If the helium fraction is indeed large ($\rm He/H \approx 2$) only part of the
accreted He/H envelope might have been ejected and steady nuclear burning 
proceeded for at least one month. This result is consistent with the 
analytical model of Kahabka (1995). Assuming an X-ray on-time of 0.1~years and
a recurrence period of 10~years, we constrain the WD mass to $\rm M_{WD} \ge 
1.36\ M_{\odot}$. U~Sco and RN in general are therefore probably SN~Ia 
progenitors (cf. Li \& van den Heuvel 1997; for a recent review on SN~Ia, 
see Livio 1999).

\section{Conclusions}
For the first time a RN (U~Sco) has been detected in a SSS X-ray phase with 
the BeppoSAX X-ray satellite $\sim$20 days after an optical outburst. This 
observation confirms the theoretical predictions that RN have a SSS X-ray 
phase (Yungelson et al. 1996; Kato 1996). He enhanced non-LTE WD atmosphere 
model spectra, with a high N/C ratio (using abundances derived from the 
ejecta) are required to fit the BeppoSAX X-ray spectrum of U Sco. This is 
evidence that the outburst of U~Sco was triggered by a TR and that the CNO 
cycle was active. From the temperature of the optically thick SSS component 
of $\rm \sim9\times 10^5\ K$, we constrain the WD to be very massive 
($\rm > 1.2\ M_{\odot}$) and consistent to be close to the CH limit.

Besides the SSS emission we observe an additional optically thin component. 
We explain this hard component as emission from a strong shocked wind from 
the WD with a mass loss rate of $\rm \dot{M}\ =\ (2.4-6.9)\times 10^{-8}\ 
M_{\odot}\ yr^{-1}\ (d/kpc)$. Such a component is consistent with the 
theoretical predictions for a WD with a mass just below the CH mass 
(Hachisu et al. 1999). According to their calculations U~Sco emerged 
from an optically thick wind phase when the BeppoSAX observation was 
performed and it cannot last longer than 20~days for a mass just below 
the CH limit.

U Sco, and therefore RN in general, can be considered to be progenitors of 
SN~Ia. The condition that the WD can grow in mass is achieved if the accreted 
and accumulated material is enriched in He and not all the envelope was 
ejected. This condition may occur if the donor star has experienced a previous
helium accretion phase, if it is somewhat evolved (a subgiant), or if 
helium rich material has been mixed into the accreted envelope.

\acknowledgements
We thank Luigi Piro for granting the BeppoSAX TOO and the BeppoSAX team 
including Milvia Capalbi for the very fast production and delivery of the 
final observation tape. We thank L. Yungelson for discussions and the 
referee M. Orio for critical comments. This research was supported in part 
by the Netherlands Organization for Scientific Research (NWO) through 
Spinoza Grant 08-0 to E.P.J. van den Heuvel. I. Negueruela is an ESA 
External Research Fellow.


\end{document}